\documentclass[prb,superscriptaddress,twocolumn,showpacs]{revtex4}
\usepackage{amsmath}
\usepackage{bm}
\usepackage{amssymb}
\usepackage{graphicx}

\begin{document}

\title{Single-electron approach for time-dependent electron transport}

\author{Shmuel Gurvitz}
\email{shmuel.gurvitz@weizmann.ac.il}

\affiliation{Department of Particle Physics and Astrophysics\\  Weizmann Institute of
Science, Rehovot 76100, Israel}
\affiliation{Institute for Advance Studies at the Hebrew University, Jerusalem 91904,
Israel}
\affiliation{ Beijing Computational Science Research Center,
Beijing 100084, China}
\date{\today}


\begin{abstract}
We develop a new approach to electron transport in mesoscopic systems by using a particular single-particle basis. Although this basis generates redundant many-particle amplitudes, it greatly simplifies the treatment. By using our method for transport of non-interacting electrons, we generalize the Landauer formula for transient currents and time-dependent potentials. The result has a very simple form and clear physical interpretation. As an example, we apply it to resonant tunneling through a quantum dot where the tunneling barriers are oscillating in time. We obtain an analytical expression for the time-dependent (ac) resonant current.  However, in the adiabatic limit this expression displays the dc current for zero bias (electron pumping).
\end{abstract}

\maketitle
\section{Introduction}

The study of quantum transport in a mesoscopic system is one of the most active areas in condensed-matter physics. Great progress has been achieved in this field following the pioneering work of Landauer \cite{land}, which implied scattering theory to electrical conduction. However, the formal scattering theory treats only the steady-state processes, whereas in many cases we need to consider  transient currents and time-dependent potentials as well. For this reason different time-dependent extensions of the Landauer formalism have been proposed \cite{you,vanl}, based on the non-equilibrium Green's function technique.

Unfortunately, the Green's function approach is more complicated for applications than the Landauer theory. For instance, the Landauer treatment avoids all complications related to the Pauli exclusion principle inside the quantum system, currying the current. Indeed, it does not use the Fermi factors, preventing accumulation of many electrons at the same quantum state during the transport. The Pauli blocking enters there only via the Fermi factors of the reservoirs. Although the Landauer results have been reproduced by using the Green's function method, an absence of Fermi factors inside the quantum system has not been very transparent, in spite of some explanations proposed by Landauer on the issue \cite{land1}.

We believe that for a better understanding of the electron transport it is desirable to study it from the
Schr\"odinger evolution of a many-electron wave function for an entire system. In this way we can determine the proper conditions when the system can be described in terms of the time-dependent single-electron wave functions. This will also show us how the Landauer formula arises in the steady-state limit ($t\to\infty$)  and what happen with the Pauli blocking inside the system. In addition, the same treatment for {\em time-dependent} potentials would result in a new  Landauer-type formula for the time-dependent case.

We therefore refer to our procedure as a single-electron  approach, although we deal with the many-electron wave function, where the Pauli ant-symmetrization is explicitly included. In fact, this approach has been partially used in earlier publications \cite{bg}, but as a phenomenological procedure. Here we present its detailed microscopic derivation, using as a generic example the electron transport through a single quantum dot. Finally, we obtain a new simple formula for the time-dependent transport. As an example for applications of this formula, we consider the time-dependent resonant current though periodically  modulated tunneling barriers, regarded as a model for electron pumping and quantum shuttle.

\section{General treatment}

\begin{figure}[h]
\includegraphics[width=7cm]{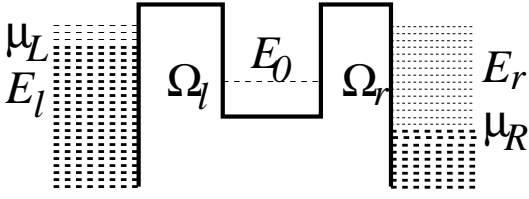}
\caption{Resonant tunneling through a single dot. $\mu_{L(R)}$ denote the Fermi levels in the left (right) lead.} \label{fig1}
\end{figure}

Consider resonant tunneling through a quantum well (quantum dot) coupled to two reservoirs, Fig.~\ref{fig1}. This system is described by the following tunneling Hamiltonian
\begin{align}
&H(t)=\sum_lE_l\hat c_l^\dagger\hat c_l+\sum_rE_r\hat c_r^\dagger \hat
c_r+E_0(t)\hat c_0^\dagger \hat c_0
\nonumber\\
&+\left(\sum_{l}\Omega_l(t)\hat
c^\dagger_l \hat c_0 +\sum_{r}\Omega_r(t)\hat c^\dagger_r \hat c_0+H.c.\right)\, ,
\label{a1}
\end{align}
where $\hat c_{l(r)}$ denotes the electron annihilation operator in the left (right) reservoir and $\hat c_{0}$ is the same inside the quantum dot. The Hamiltonian $H$ is {\em time-dependent}. It is reflected in time-dependence of the energy level $E_0(t)$ and the tunneling couplings $\Omega_{l,r}(t)$, Eq.~(\ref{a1}). The total number of electrons ($N$) in the system is conserved. (In the following we consider the limit of $N\to\infty$).

The total many-body wave function, $|\Psi (t)\rangle$ is obtained from the time-dependent Schr\"odinger equation
\begin{align}
i\partial_t|\Psi (t)\rangle =H(t)|\Psi (t)\rangle
\label{a5}
\end{align}
with the initial condition
\begin{align}
|\Psi(0)\rangle =\prod_{k}\hat c_{k}^\dagger |0\rangle\, ,
\label{a2}
\end{align}
where the index $k$ denotes the initially occupied states of the left and right reservoirs. (If each of the reservoirs is initially at zero temperature, then $k\in \{E_{l}\le \mu_L,E_{r}
\le \mu_R \}$).

In order to solve Eq.~(\ref{a5}) we use an ansatz for the total wave function $|\Psi (t)\rangle$, by taking it as a (Slater) product of single-electron wave functions,
\begin{align}
|\Psi(t)\rangle =\prod_k\hat\Phi_{}^{(k)}(t)|0\rangle\,
\label{a3}
\end{align}
where
\begin{align}
\hat\Phi_{}^{(k)}(t)=\sum_{l}b_{l}^{(k)}(t)\,\hat c_{l}^\dagger +b_{0}^{(k)}(t)\,\hat c_0^\dagger+\sum_{r}b_{r}^{(k)}(t)\,\hat c_{r}^\dagger
\label{a4}
\end{align}
Note that this wave function $|\Psi(t)\rangle$ contains many redundant multi-particles components, like $b_0^{(k_1)}(t)b_0^{(k_2)}(t)\hat c_0^{\dagger 2}|0\rangle$, which has zero contribution. Nevertheless, it would be very useful to keep these components explicitly, as will be clear from the following derivations.

Let us substitute (\ref{a3}), (\ref{a4}) into the Schr\"odinger equation (\ref{a5}). We find
\begin{align}
&i\partial_t|\Psi (t)\rangle=
\sum_k\prod_{k'<k}\hat\Phi_{}^{(k')}(t)\,\big[i\,
\partial_t\hat\Phi_{}^{(k)}(t)\big]
\prod_{k''>k}\hat\Phi_{}^{(k'')}(t)|0\rangle
\nonumber\\
&=
\sum_k\prod_{k'<k}\hat\Phi_{}^{(k')}(t)\,\Big[
H,\hat\Phi_{}^{(k)}(t)\Big]
\prod_{k''>k}\hat\Phi_{}^{(k'')}(t)|0\rangle
\label{app4}
\end{align}
Then it follows from this equation that
\begin{align}
i\, \partial_t\hat\Phi_{}^{(k)}(t)=\Big[
H,\hat\Phi_{}^{(k)}(t)\Big]
\end{align}
where the commutator can be written explicitly as
\begin{align}
&\Big[H(t),\hat\Phi^{(k)}(t)\Big]=\sum_{l}
\big[E_{l}b_{l}^{(k)}(t)
+\Omega_l(t)b_{0}^{(k)}(t)\big]
\,\hat c_{l}^\dagger
\nonumber\\
&+\big[E_0(t)b_{0}^{(k)}(t)+\sum_l\Omega_l(t)\,b_{l}^{(k)}(t)
+\sum_r\Omega_r(t)\,b_{r}^{(k)}(t)
\big]\,\hat c_0^\dagger\nonumber\\[5pt]
&+\sum_{r}\big[E_rb_{r}^{(k)}(t)+\Omega_r(t)b_{0}^{(k)}(t)\big]
\,\hat c_{r}^\dagger
\label{app4p}
\end{align}

Then using Eq.~(\ref{app4p}) we find  the following system of linear equations for amplitudes $b_{}^{(k)}(t)$
\begin{subequations}
\label{a6}
\begin{align}
i\dot {b}_{l}^{(k)}(t)&=E_l\,b_{l}^{(k)}(t)+\Omega_l(t)\,b_{0}^{(k)}(t)
\label{a6a}\\
i\dot {b}_{0}^{(k)}(t)&=E_0(t)\,b_{0}^{(k)}(t)+\sum_l\Omega_l(t)\,b_{l}^{(k)}(t)
\nonumber\\
&~~~~~~~~~~~~~~~~~~~~~~~~~~~
+\sum_r\Omega_r(t)\,b_{r}^{(k)}(t)\label{a6b}\\
i\dot {b}_{r}^{(k)}(t)&=E_r\,b_{r}^{(k)}(t)+\Omega_r(t)\,b_{0}^{(k)}(t)\, ,\label{a6c}
\end{align}
\end{subequations}
supplemented with the initial condition $b_{l}^{(k)}(0)=\delta_{kl}$, $b_{r}^{(k)}(0)=\delta_{kr}$ and $b_{0}^{(k)}(0)=0$. Note that the amplitudes corresponding to different initial single-electron states $k$ are {\em decoupled} in Eqs.~(\ref{a6}). This is a great advantage of the single-electron approach with respect to alternative treatments. Unfortunately, in the case of interaction (third- and higher-order terms in creation and annihilation operators), the amplitudes with different $k$ are not decoupled in general. Nevertheless, in some cases of interaction, the method can be applied as well. For instance, it takes place when the quantum dot is coupled capacitively to the electrostatic qubit, such that the interaction term commutes with the qubit's Hamiltonian (so-called ``partial decoherence'' \cite{pd}). However, it is not considered in this paper.

Equations~(\ref{a6a}) and (\ref{a6c}) can be solved explicitly thus obtaining
\begin{align}
b_{\alpha}^{(k)}(t)=e^{-iE_{\alpha}t}\Big[\delta_{k\alpha}
-\int\limits_{0}^t
i\,\Omega_{\alpha}(t') b_{0}^{(k)}(t')e^{iE_{\alpha}t'}dt'\Big]
\label{a120}
\end{align}
where $\alpha =l,r$. Substituting $b_{l,r}^{(k)}(t)$ from Eq.~(\ref{a120}) into
Eq.~(\ref{a6b}) we can rewrite it as
\begin{align}
&i\dot {b}_{0}^{(k)}(t)=E_0(t)\,b_{0}^{(k)}(t)+
\Omega_k(t)e^{-iE_kt}\nonumber\\
&-i\sum_l\Omega_l(t)\int\limits_0^t\Omega_l(t')b_0^{(k)}(t')
e^{iE_l(t'-t)}dt'\nonumber\\
&-i\sum_r\Omega_r(t)\int\limits_0^t\Omega_r(t')b_0^{(k)}(t')
e^{iE_r(t'-t)}dt'
\label{a6bb}
\end{align}
Note that $\sum_{l,r}$ extend over {\em all} reservoir states ($E_{l,r}$) without any Pauli principle restrictions.

Consider the continuous limit. Then we replace
$\sum_{l,r}\to\int\varrho_{L,R}dE_{l,r}$, where $\varrho_{L,R}$ are the density of states in the left and the right lead. We assume that $\varrho_{L,R}$ are energy independent and so the tunneling couplings,  $\Omega_{l,r}\to\Omega_{L,R}$ (so-called wide band limit). Using $\int_{-\infty}^\infty e^{iE_{l,r}(t'-t)}dE_{l,r}=2\pi\delta (t'-t)$, we find that Eq.~(\ref{a6bb}) becomes
\begin{align}
\dot {b}_{0}^{(k)}(t)=\left(-iE_0(t)-{\Gamma(t)\over2}\right)
\,b_{0}^{(k)}(t)-i\,\Omega_k(t) e^{-iE_kt}
\label{a6f}
\end{align}
where $\Gamma(t)=\Gamma_L(t)+\Gamma_R(t)$ is a total (time-dependent) width of the energy level $E_0$, while $\Gamma_{L,R}(t)=2\pi\Omega_{L,R}^2(t)\varrho_{L,R}$ are corresponding (time-dependent) partial widths.  Note that the coupling $\Omega_{L,R}$ scales as $1/\sqrt{\bar L}$ with a reservoir size $\bar L$, whereas the density of states $\varrho\propto{\bar L}$. As a results the product $\Omega_{L,R}^2\varrho_{L,R}$ remains finite in a continuous limit ($\bar L\to\infty$).
Solving Eq.~(\ref{a6f}) we obtain
\begin{align}
b_0^{(k)}(t)=-i\,e^{-i{\cal E}_0(t)t} \int\limits_0^t
\Omega_k(t')e^{i[{\cal E}_0(t')-E_k]t'}dt'
\label{a6fin}
\end{align}
with
\begin{align}
{\cal E}_0(t)={1\over t}\int\limits_0^t \left [E_0(t')-i{\Gamma (t')\over2}\right]dt'
\label{calen}
\end{align}
In the case of time-independent Hamiltonian, ${\cal E}_0(t)=E_0-i\Gamma/2$ we obtain from Eq.~(\ref{a6fin})
\begin{align}
b_{0}^{(k)}(t)={\Omega_k\,e^{-iE_{k}t}\over E_{k}-E_0+i{\Gamma\over2}}\big[ 1-e^{i(E_{k}-E_0)t-{\Gamma\over2}t}\big]
\label{a12}
\end{align}
Thus in the asymptotic limit $t\to\infty$ the amplitude $b_{0}^{(k)}(t)$ becomes a Lorentzian centered at $E_0$.

\subsection{Occupation of the dot.}

Using Eqs.~(\ref{a120}) and (\ref{a6fin}) we obtain the total  wave-function, Eqs.~(\ref{a3}), (\ref{a4}) which allows us to evaluate all observables. We start with (average) charges in the right (left) reservoir, $Q_{R(L)}(t)$, and inside the quantum dot, $Q_0(t)$ given by
\begin{align}
&Q_{L(R)}(t)
=\langle\Psi(t)|\sum_{l(r)}\hat n_{L(R)}^{}
|\Psi(t)\rangle\\
&Q_{0}(t)=\langle\Psi(t)|\hat n_0^{}
|\Psi(t)\rangle
\label{charge}
\end{align}
where $\hat n_{L(R)}^{}=\sum_{l(r)}\hat c^\dagger_{l(r)}\hat c_{l(r)}^{}$, and $\hat n_{0}^{}=\hat c^\dagger_{0}\hat c_{0}^{}$, and $|\Psi(t)\rangle$ is given by Eq.~(\ref{a3}). Consider for example $Q_0(t)$. We can write it explicitly as
\begin{align}
Q_0(t)=\langle 0|\hat\Phi_{}^{(N)\dagger }\cdots\hat\Phi_{}^{(1)\dagger}
\hat c^\dagger_{0}\hat c_{0}^{}\hat\Phi_{}^{(1)}\cdots\hat\Phi_{}^{(N)}
|0\rangle
\end{align}
where we enumerated the occupied state in the reservoirs as $k=\{1,2,\ldots N\}$. Using Eq.~(\ref{a4}) we can rewrite $Q_0(t)$ as
\begin{align}
&Q_0(t)=\langle 0|\hat\Phi_{}^{(N)\dagger }\cdots\hat\Phi_{}^{(2)\dagger}
\hat\Phi_{}^{(2)}\cdots\hat\Phi_{}^{(N)}
|0\rangle\,|b_0^{(1)}(t)|^2\nonumber\\
&+\langle 0|\hat\Phi_{}^{(N)\dagger }\cdots\hat c^\dagger_{0}\hat\Phi_{}^{(1)\dagger}
\hat\Phi_{}^{(1)}\hat c_{0}^{}\cdots\hat\Phi_{}^{(N)}
|0\rangle
\end{align}
Note that $\langle 0|\hat\Phi_{}^{(N)\dagger }\cdots\hat\Phi_{}^{(2)\dagger}
\hat\Phi_{}^{(2)}\cdots\hat\Phi_{}^{(N)}
|0\rangle =1$, since it is a normalization of the ($N-1$) electron wave function.

We can continue with this procedure for evaluating $Q_0(t)$ and the same for $Q_{L(R)}(t)$, finally obtaining
\begin{align}
&Q_{0}(t)=\sum_k|b_{0}^{(k)}(t)|^2\nonumber\\
&Q_{L(R)}(t)
=\sum_k\sum_{l(r)}|b_{l(r)}^{(k)}(t)|^2
\label{charge1}
\end{align}
Here the sum over $l,r$ is extended over all states (occupied or not) of the leads, whereas the sum over $k$ includes only the initially occupied levels of the left and right leads. Note that the total charge is conserved in time, so that $Q_L(t)+Q_R(t)+Q_0(t)=N$.

Consider the limit $N\to \infty$ and assume that the initial electron distributions in the left (right) lead is given by the Fermi function $f_{L(R)}(E)$. Then sum over $k$ is replaced by  integral $\sum_k\to\int\limits_{-\infty}^{\infty}f_{L(R)}(E)\varrho_{L(R)}^{}dE$. As a result
\begin{align}
&Q_{0}^{}(t)=\int\limits_{-\infty}^{\infty}|b_{0}^{(\bar l)}(t)|^2
f_L(E_{\bar l})\varrho_L^{}dE_{\bar l}\nonumber\\
&+\int\limits_{-\infty}^{\infty}|b_{0}^{(\bar r)}(t)|^2f_R(E_{\bar r})
\varrho_R^{}dE_{\bar r}
\equiv Q_{0}^{(L)}(t)
+Q_{0}^{(R)}(t)
\label{charge1c}
\end{align}
where $\bar l(\bar r)$ denote the initially occupied states in the left (right) lead. Respectively $Q_{0}^{(L,R)}(t)$ denote the charge coming to the dot from the left (right) lead.

Using Eq.~(\ref{a6fin}) we can rewrite Eq.~(\ref{charge1c}) as
\begin{align}
Q_0^{(\alpha)}(t)=\int\limits_{-\infty}^{\infty}n^{(\alpha)}_{}(E,t)
f_{\alpha}^{}(E)dE
\label{oc12}
\end{align}
where $\alpha =L,R$ and
\begin{align}
n^{(\alpha)}_{}(E,t)=\left |\int\limits_0^t\Omega_{\alpha}^{}(t')
e^{i[{\cal E}_0(t')-E]t'-i{\cal E}_0(t)t}dt'\right|^2\varrho_{\alpha}^{}
\label{oc11p}
\end{align}
For the time-independent Hamiltonian, $E_0(t)=E_0$ and $\Omega_{\alpha}^{} (t)=\Omega_{\alpha}^{}$ one easily finds from Eq.~(\ref{oc11p})
\begin{align}
n^{(\alpha )}(E,t)={1-2\cos (E-E_0)t\,e^{-{\Gamma\over2}t}+e^{-\Gamma t}\over (E-E_0)^2+{\Gamma^2\over4}}\,\,{\Gamma_{\alpha}\over 2\pi}
\label{oc11}
\end{align}

Let us take the  reservoirs at zero temperature, which corresponds to $f_{L,R}(E)=\theta (\mu_{L,R}-E)$, in  Eq.~(\ref{oc12}). Then one easily obtains for average charge of the dot in the steady-state limit, $\bar q=Q_0(t\to\infty)$
\begin{align}
\bar q={\Gamma_L\over\Gamma}\Big(1-{1\over \pi}\arg\nu_L^{}\Big)+{\Gamma_R\over\Gamma}
\Big(1-{1\over \pi}\arg\nu_R^{}\Big)
\label{ocst}
\end{align}
where $\nu_{L,R}^{}=\mu_{L,R}^{}-E_0+i{\Gamma\over2}$. As expected, the dot is fully occupied only for $\mu_{L,R}^{}\to\infty$.
If $\mu_R^{}=\mu_L^{}=E_0^{}$, the average occupation of the dot is $1/2$.

\subsection{Currents}

The average current in the left (right) lead is defined as $I_{L(R)}(t)=\dot Q_{L(R)}(t)$. Using Eqs.~(\ref{charge1}) and (\ref{charge1c}) we can write
\begin{subequations}
\label{rescur2}
\begin{align}
&I_{L}(t)=\dot Q_{L}^{(L)}(t)+\dot Q_{L}^{(R)}(t)
\equiv I_{L\to L}(t)+I_{R\to L}(t)
\label{rescur2a}\\
&I_{R}(t)=\dot Q_{R}^{(L)}(t)+\dot Q_{R}^{(R)}(t)
\equiv I_{L\to R}(t)+I_{R\to R}(t)
\label{rescur2b}
\end{align}
\end{subequations}
where $I_{L\to R}(t)={d\over dt}\sum_{\bar l, r}|b_r^{(\bar l)}(t)|^2$ is the right-lead current originated by electrons initially occupied in the left lead and $I_{R\to R}(t)={d\over dt}\sum_{\bar r, r}|b_r^{(\bar r)}(t)|^2$ is the same for electrons initially occupied in the right lead. Similarly
$I_{R\to L}(t)={d\over dt}\sum_{\bar r,l}|b_l^{(\bar r)}(t)|^2$ and $I_{L\to L}(t)={d\over dt}\sum_{\bar l,l}|b_l^{(\bar l)}(t)|^2$ denote the same components of the left-lead current.
Consider for instance the current from the left to right lead, $I_{L\to R}(t)$. Using Eqs.~(\ref{a6}) and (\ref{a120}) it can be written as
\begin{align}
&I_{L\to R}(t)=\sum_{\bar l}\,{\rm Re}\,\int\limits_0^t dt'\int\limits_{-\infty}^\infty 2\,\Omega_R^{}(t)\Omega_R^{}(t')\, b_0^{(\bar l)}(t) b_0^{(\bar l)*}(t')\nonumber\\
&\times \varrho_R
e^{iE_{r}(t-t')}dE_r=
\Gamma_R^{}(t)\int\limits_{-\infty}^{\infty}|b_{0}^{(\bar l)}(t)|^2
f_L(E_{\bar l})\varrho_LdE_{\bar l}\nonumber\\
&~~~~~~~~~~~=\Gamma_R^{}(t)Q_0^{(L)}(t)
\label{rescur1}
\end{align}
where $Q_0^{(L)}(t)$ is given by Eq.~(\ref{oc12}).

In order to evaluate the second component of the right-lead current, namely $I_{R\to R}(t)=\dot Q_R^{(R)}(t)$, in the most simple way, we imply the conservation of probability for a single-electron wave function,
\begin{align}
\sum_l |{b}_{l}^{(k)}(t)|^2+|{b}_{0}^{(k)}(t)|^2+
\sum_r |{b}_{r}^{(k)}(t)|^2=1
\label{probcon}
\end{align}
We obtain
\begin{align}
&I_{R\to R}(t)=-{d\over dt}\sum_{\bar r}\sum_l |b_{l}^{(\bar r)}(t)|^2-{d\over dt}\sum_{\bar r}|b_{0}^{(\bar r)}(t)|^2\nonumber\\
&=-\int\limits_{-\infty}^{\infty}\Gamma_L^{}(t)|b_{0}^{(\bar r)}(t)|^2 f_R(E_{\bar r})\varrho_RdE_{\bar r}
-{d\over dt}Q_0^{(R)}(t)\nonumber\\
&=-\Gamma_L^{}(t)Q_0^{(R)}(t)-\dot Q_0^{(R)}(t)
\label{probcon1}
\end{align}
where $Q_0^{(R)}(t)$ is given by Eq.~(\ref{oc12}). Finally we find for the right-lead current, Eq.~(\ref{rescur2b})
\begin{align}
I_{R}(t)=\Gamma_R^{}(t)Q_0^{(L)}(t)-\Gamma_L^{}(t)Q_0^{(R)}(t)
-\dot Q_0^{(R)}(t)
\label{a11}
\end{align}
Using Eq.~(\ref{oc12}) this expression can be rewritten as
\begin{align}
&I_{R}(t)=\int\limits_{-\infty}^\infty\Big[n^{(L)}_{}(E,t)
\Gamma_R^{}(t)f_L(E)\nonumber\\
&- n^{(R)}_{}(E,t)
\Gamma_L^{}(t)f_R(E)
-\dot n^{(R)}_{}(E,t)f_{R}^{}(E)\Big]dE
\label{rcurr}
\end{align}

Each term of Eq.~(\ref{a11}) has a simple physical meaning. The first one describes the current from the left to right leads through the dot with the incoming rate $\Gamma_L^{}(t)$ (incorporated in $Q_0^{(L)}(t)$, Eqs.~(\ref{oc12}), (\ref{oc11p})), and the outgoing rate $\Gamma_R^{}(t)$. The second term of Eq.~(\ref{a11}) describes the reverse process, from the right to the left reservoir. The last term, the so-called ``displacement current''\cite{you}, describes the time-dependent occupation of the dot (which was empty at $t=0$) by electrons coming from the left lead. This temporary process takes place until the dot reaches its steady state.

In general, the circuit current is $I(t)=c_R^{} I_R^{}(t)-c_L^{} I_L^{}(t)$, where the coefficients $c_{L,R}^{}$ with $c_L^{}+c_R^{}=1$ depend on the junction capacities \cite{bb}. Here the left-lead current $I_L(t)$ is given by the same Eq.~(\ref{rcurr}) with $L\leftrightarrow R$. Using the current conservation, $I_L(t)+I_R(t)+\dot Q_0(t)=0$ and Eqs.~(\ref{charge1c}),  (\ref{rcurr}) we can write the circuit-current explicitly as
\begin{align}
&I(t)=\int\limits_{-\infty}^\infty\Big[n^{(L)}_{}(E,t)
\Gamma_R^{}(t)f_L(E)- n^{(R)}_{}(E,t)
\Gamma_L^{}(t)f_R(E)\nonumber\\
&
-c_R^{}\dot n^{(R)}_{}(E,t)f_{R}^{}(E)+c_L^{}\dot n^{(L)}_{}(E,t)f_{L}^{}(E)\Big]dE
\label{curr1}
\end{align}
The first two terms of Eq.~(\ref{curr1}) look similar to the Landauer formula. The remaining terms represent the displacement current. In the case of time-independent Hamiltonian, the displacement current vanishes in the steady-state limit, $t\to\infty$, Eq.~(\ref{oc11}). Then we find for the steady-state current, $\bar I=I(t\to\infty)$
\begin{align}
\bar I=\int\limits_{-\infty}^{\infty}{\Gamma_L\Gamma_R\over (E-E_0)^2+{\Gamma^2\over 4}}\big[f_L(E)-f_R(E)\big]{dE\over 2\pi}
\label{land}
\end{align}
This result coincides with the Landauer formula for an electron current through a single level. Therefore, Eq.~(\ref{curr1}) can be considered an extension of the Landauer formula for the time-dependent case.

Let us evaluate the time-dependent linear response (conductance), $G(E,t)$, for the reservoirs at zero temperature. We can write it as
\begin{align}
G(E,t)&={1\over2}[n^{(L)}_{}(E,t)\Gamma_R^{}(t)
+n^{(R)}_{}(E,t)\Gamma_L^{}(t)
\nonumber\\
&+
c_L^{}\dot n^{(L)}_{}(E,t)+c_R^{}\dot n^{(R)}_{}(E,t)]
\label{lires}
\end{align}
where $E$ is the Fermi energy. The last two terms represents a contribution from the displacement current.

Consider the initial condition corresponding to the empty dot.
In this case $n^{(L,R)}_{}(E,t)$ is given by Eq.~(\ref{oc11}).
For a symmetric case, $\Gamma_L=\Gamma_R=\Gamma/2$ and $c_L^{}=c_R^{}=1/2$, we easily obtain
\begin{align}
G(E,t)={\Gamma^2\over 2\pi ( 4E^2+\Gamma^2)}\left[1-\Big(\cos Et-{2E\over\Gamma}\sin Et\Big)e^{-{\Gamma t\over2}}\right]
\end{align}
Here we chose a scale where $E_0=0$. One finds that in the asymptotic limit, $t\to\infty$, the conductance coincides with that given by Eq.~(\ref{land}). However, for a finite $t$ it displays damped oscillations. One finds from Eqs.~(\ref{oc11}), (\ref{curr1}) that the contribution from the displaced current is quite important in this region.

\section{Oscillating barriers}

Equation (\ref{curr1}) is valid for any time-dependence (random or regular) of tunneling couplings and energy levels in the Hamiltonian (\ref{a1}). As an example, we consider periodic
oscillations of tunneling couplings
\begin{align}
\Omega_{L,R}(t)=\Omega[1\pm \xi\sin (\omega t)]
\end{align}
This can be a model for electron pumping \cite{mosk} or quantum shuttle \cite{shekh}. Respectively, the tunneling widths are
\begin{align}
\Gamma_{L,R}(t)={\Gamma\over2}\,\Big[1+{\xi^2\over2}\pm 2\xi\sin (\omega t)-{\xi^2\over2}\cos (2\omega t)\Big]
\label{pwidth}
\end{align}
where $\Gamma =4\pi\Omega^2\varrho$. (For simplicity we consider  $\varrho_L^{}=\varrho_R^{}=\varrho$). The total width is therefore
\begin{align}
\Gamma (t)=\Gamma_L(t)+\Gamma_R(t)=\Gamma \Big[1+{\xi^2\over2}-{\xi^2\over2}\cos (2\omega t)\Big]
\label{twidth}
\end{align}

Substituting (\ref{twidth}) into Eq.~(\ref{calen}) we obtain
\begin{align}
{\cal E}_0(t)t=E_0t-i{\Gamma \over2}\left[ \left( 1+{\xi^2\over2}\right)t-{\xi^2\over 4\omega}\sin (2\omega t)\right ]
\end{align}
Using this expression we evaluate $n^{(L,R)}(E,t)$, Eq.~(\ref{oc11p}). In order to obtain simple analytical expressions we keep only the terms up to $\xi^2$. Then
\begin{align}
n^{(\alpha)}_{}
=&{\Gamma\over4\pi}\left|\int\limits_0^t\Big\{1 \pm\xi\sin(\omega t')+{\Gamma \xi^2\over 8\omega}[\sin (2\omega t)-\sin (2\omega t')]\Big\}\right.
\nonumber\\
&\left.\times
e^{{\tilde\Gamma\over2}(t'-t)-iEt'}
dt'\right|^2
\label{curr2}
\end{align}
where $\alpha =L,R$, corresponding to $\pm\xi$ and $\tilde\Gamma =\Gamma\big(1+{\xi^2\over2}\big)$ is a renormalized total width.
Here again $E_0=0$.

One can perform the integration in Eq.~(\ref{curr2}) analytically, thus obtaining
\begin{align}
&n^{(L,R)}_{}(E,t)=\left|\left[1+{\Gamma \xi^2\over 8\omega}\sin (2\omega t)\right]f_0^{}(E,t)\pm\xi f_1^{}(E,t)\right.\nonumber\\
&\left.~~~~~~~~~~~~~~~~~~~~~~~~~~~~~~~~~~~
-{\Gamma\xi^2\over 8\omega}f_2^{}(E,t)
\right|^2{\Gamma\over4\pi}
\label{nal}
\end{align}
where
\begin{align}
&f_0^{}={e^{-iEt}-e^{-{\tilde\Gamma\over 2}t}\over {\tilde\Gamma\over2}-i E}~~~{\rm and}\nonumber\\
&f_j^{}={i\over 2}
\left[{e^{-i(E+j\omega )t}-e^{-{\tilde\Gamma\over 2}t}\over {\tilde\Gamma\over2}-i (E+j\,\omega)}
-{e^{-i(E-j\omega )t}-e^{-{\tilde\Gamma\over 2}t}\over {\tilde\Gamma\over2}-i (E-j\,\omega)}\right]
\label{nal1}
\end{align}
with $j=1,2$.
Substituting this result into Eq.~(\ref{curr1}), we obtain the time-dependent current, $I(t)$.

Let us consider the initial state of the reservoirs corresponding  to zero temperature and zero bias, $\mu_L=\mu_R=\mu$, and take for simplicity $c_L^{}=c_R^{}=1/2$. In this case Eq.~(\ref{curr1}) reads \begin{align}
&I(t)=\int\limits_{-\infty}^\mu \Big[n^{(L)}_{}(E,t)
\Gamma_R^{}(t)- n^{(R)}_{}(E,t)
\Gamma_L^{}(t)\nonumber\\
&+{1\over2}\big(\dot n^{(L)}_{}(E,t)-\dot n^{(R)}_{}(E,t)\big)\Big]dE
\label{curr3}
\end{align}
where $\Gamma_{L,R}^{}(t)$ and $n^{(L,R)}_{}(t)$ are given by Eqs.~(\ref{pwidth}), (\ref{nal}) and (\ref{nal1}). Consider the limit of $t\to\infty$ when the terms $\propto \exp (-\tilde\Gamma t/2)$ in (\ref{nal1}) vanish. Then we obtain the following result for the time-dependent total current in the asymptotic  region
\begin{align}
I(t)={\Gamma \xi\over 8\pi}[A(\mu,\omega)\cos (\omega  t)+B(\mu,\omega)\sin (\omega t)+O[\xi^3]
\label{ascur1}
\end{align}
where
\begin{align}
&A(\mu,\omega)=\ln {4(\mu-\omega)^2+\Gamma^2\over 4(\mu+\omega)^2+\Gamma^2},~~
B(\mu,\omega)=-2 \tan^{-1}\left(\frac{2
\mu}{\Gamma }\right)\nonumber\\
&+\tan ^{-1}\left(\frac{2(\mu-\omega )}{\Gamma}\right)+\tan^{-1}\left(\frac{2
(\mu+\omega )}{\Gamma}\right)
\end{align}

Consider the adiabatic limit of $I(t)$, by expanding Eq.~(\ref{ascur1}) in powers of $\omega$. One finds
\begin{align}
I(t)=-{2\mu\,\Gamma\,\xi\,\omega\over \pi(4\mu^2+\Gamma^2)}+O\big[(\omega/\Gamma)^3\big]
\label{adiab}
\end{align}
This result is very remarkable, since it displays no time-dependence in the resonant current up to the terms of order $\omega^3$. It implies that in the adiabatic limit there exists the dc current for zero bias voltage, generated by barriers oscillations. This would correspond to the so-called ``electron pumping'' widely discussed in the literature\cite{mosk}.

\begin{figure}[h]
\includegraphics[width=7cm]{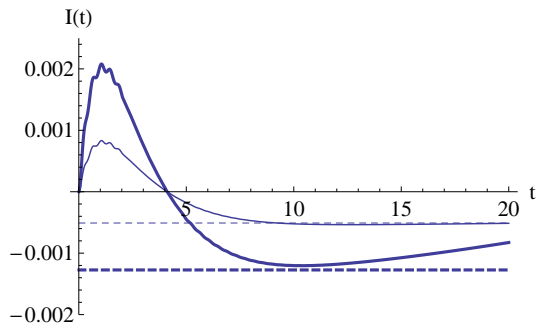}
\caption{Resonant current trough quantum dot for oscillating barriers (electron pumping) for two values of the oscillation frequency: $\omega=0.02 \Gamma$ (thin line) and $\omega =0.05 \Gamma$ (thick line).} \label{fig2}
\end{figure}

In Fig.~\ref{fig2} we show the time-dependent current (in units of $\Gamma$), given by Eq.~(\ref{curr3}) (solid lines) and its adiabatic limit, Eq.~(\ref{adiab}) (dashed lines) for $\xi=0.2$, $\mu=0.25 \Gamma$ and two values of the oscillation frequency: $\omega =0.02 \Gamma$ (thin) and $\omega =0.05 \Gamma$ (thick)
One finds from this figure that the adiabatic limit works very well when $\omega/\Gamma$ becomes small enough. Indeed $\omega =.02 \Gamma$ one find the dc current (the electron pumping) after the transient period due to the initial conditions corresponding to  empty dot. Nevertheless for larger times. not shown in Fig.~\ref{fig2}, the current will display oscillations, as follows from Eq.~(\ref{ascur1}).

\section{Summary}

In this paper we present a detailed microscopic derivation of the single-electron approach to electron transport in time-dependent potentials, based on the single-electron ansatz for the many-electron function. The final expression for the time-dependent resonant current is very simple for application. It also includes the displacement current, which is in particular important  for time-dependent potentials. As an example for application of our result, we obtain a simple analytical expression for the resonant tunneling through periodically modulated barriers,  reproducing the adiabatic and the non-adiabatic limit as well. In the adiabatic limit it displays the dc current, even in the case of zero bias voltage (electron pumping).

The results obtained in the paper are restricted to non-interacting electrons. The quantum mechanical treatment of interaction always represents a formidable problem, even if only vibrational modes of the quantum dot in Fig.~\ref{fig1} are included \cite{vaclav}. However at some condition, the single-electron ansatz can be applied in the case of interaction (as for partial decoherence \cite{pd}), thus providing a very simple treatments of complicated physical problems. In addition, our approach can be used for derivation of master equations for the reduces density matrix, describing electron transport for any bias voltages and  temperatures. The resulting equations, which are non-Markovian in general, would be very useful to understand the transition between quantum and classical descriptions. All these issues will be discussed in separate works.

\begin{acknowledgements}
The hospitality of the Kavli Institute for Theoretical Physics
China, CAS, Beijing 100190, China, where a part of this work had been done, is
gratefully acknowledged. This work was supported by the Israel Science Foundation
under grant No.\ 711091. The author is thankful to A. Aharony, O.Entin-Wohlman,
T. Brandes and G. Schaller for very useful discussions.
\end{acknowledgements}

\end{document}